\documentclass[twocolumn,preprintnumbers,amssymb,amsmath,superscriptaddress,letterpaper,nofootinbib]{revtex4}[12pt]
\usepackage{times,graphics}
\usepackage[colorlinks]{hyperref}

\usepackage{graphicx}
\usepackage{dcolumn}
\usepackage{bm}
\usepackage{natbib}
\usepackage{pstricks}
\usepackage{epsfig}
\usepackage{epstopdf}
\usepackage{multirow}
\usepackage{amsmath}
\usepackage{lipsum}

\newcommand{\be}{\begin{equation}}
\newcommand{\ee}{\end{equation}}
\newcommand{\bea}{\begin{eqnarray}}
\newcommand{\eea}{\end{eqnarray}}

\makeatletter
\newcommand*{\shifttext}[2]{%
	\settowidth{\@tempdima}{#2}%
	\makebox[\@tempdima]{\hspace*{#1}#2}%
}
\makeatother

\begin{document}

\title{Dark Energy as a Critical Phenomenon: a Resolution for Hubble Tension}

\author{Abdolali Banihashemi}
\email{a\_banihashemi@sbu.ac.ir}
\affiliation{Department of Physics, Shahid Beheshti University, 1983969411,  Tehran, Iran}

\author{Nima Khosravi}
\email{n-khosravi@sbu.ac.ir}
\affiliation{Department of Physics, Shahid Beheshti University, 1983969411,  Tehran, Iran}

\author{Arman Shafieloo}
\email{shafieloo@kasi.re.kr}
\affiliation{Korea Astronomy and Space Science Institute, Daejeon 34055, Korea and\\University of Science and Technology, Daejeon 34113, Korea}

\date{\today}

\begin{abstract}
We propose a dark energy model based on the physics of critical phenomena which is consistent with both the Planck's CMB and the Riess et al.'s local Hubble measurements. In this model the dark energy density behaves like the magnetization of the Ising model. This means the dark energy is an emergent phenomenon and we named it critically emergent dark energy model, CEDE. In CEDE, dark energy emerges at a transition redshift, $z_c$, corresponding to the critical temperature in critical phenomena. Combining the Planck CMB data and local measurement of the Hubble constant from Riess et al. (2019) we find statistically significant support for this transition with respect to the case of very early transition that represents effectively the cosmological constant. This is understandable since CEDE model naturally prefers larger values of Hubble constant consistent with local measurements. Since CEDE prefers a non-trivial transition when we consider both high redshift Planck CMB data and local Hubble constant measurements, we conclude that $H_0$ tension can be a hint for the substructure of the dark energy as a well-studied properties of critical phenomena.

\end{abstract}
\maketitle
\section{Introduction:}
Riess et al. \cite{Riess:2019cxk,Riess:2018byc,Riess:2016jrr} have devoted efforts to measure and constrain the present Hubble parameter and their most recent result shows $H_0=74.03\pm 1.42$ km/s/Mpc by the analysis of the Hubble Space Telescope observations using 70 long-period Cepheids in the Large Magellanic Cloud. This observation is in $4.4\sigma$ tension with the prediction of CMB observations by Planck satellite \cite{Aghanim:2018eyx}. This tension can be due to systematics, as speculated e.g. by \cite{Efstathiou:2020wxn}, but its chance has become less and less thanks to the independent local measurements of $H_0$ e.g. based on Type II supernovae \cite{deJaeger:2020zpb} ($H_0=75.8^{+5.2}_{-4.9}$ km/s/Mpc) or based on strong gravitational lensing effects on quasar systems, H0LiCOW, ($H_0=73.5^{+1.7}_{-1.8}$ km/s/Mpc) \cite{Wong:2019kwg,Liao:2019qoc,Liao:2020zko} or based on a calibration of the Tip of the Red Giant Branch, TRGB, ($H_0=69.8\pm1.9$ km/s/Mpc) \cite{Freedman:2019jwv} or based on calibration of the SN Ia luminosity using highly-evolved low-mass stars, Miras, ($H_0=73.6\pm3.9$ km/s/Mpc) \cite{Huang:2019yhh} or based on calibration of the type Ia supernovae with surface brightness fluctuations, SBF, ($H_0=76.5\pm4.0$ km/s/Mpc) \cite{Khetan:2020hmh} or by using geometric distance measurements to megamaser-hosting galaxies, MCP, ($H_0=74.8\pm3.1$ km/s/Mpc) \cite{Pesce:2020xfe}.  In addition it is worth to mention that the $\Lambda$CDM model suffers from some other (mild) tensions e.g. $S_8$ \cite{Heymans:2020gsg}, low/high $\ell$ \cite{Addison:2015wyg} and CMB spatial anomalies \cite{Eriksen:2003db,Vielva:2003et,deOliveira-Costa:2003utu}. Nevertheless, $H_0$ tension seems the most robust one supported by many observations. In near future the situation will be clear and if this tension be real then it is very important to know if there is any theoretical explanation for it.

There are many different theoretical attempts to address the $H_0$ tension. Dynamical dark energy models \cite{Yang:2018prh,Keeley:2019esp,DiValentino:2020naf,Benaoum:2020qsi,Yang:2020myd,Zhao:2017cud,Sahni:2014ooa} and gravity theories in which gravity changes with redshift \cite{Khosravi:2017hfi,Yan:2019gbw,Frusciante:2019puu,Braglia:2020iik,Raveri:2019mxg} are examples of the late universe modifications. On the other hand, assuming an early dark energy phase before recombination in order to decrease the sound horizon \cite{Poulin:2018cxd,Karwal:2016vyq,Pettorino:2013ia} or non-standard recombination scenarios \cite{Chiang:2018xpn,Jedamzik:2020krr,Sekiguchi:2020teg,Bose:2020cjb} are the main topics of the early universe solutions. There have been though some serious doubts on how these early dark energy models can practically resolve the Hubble tension \cite{Hill:2020osr}. 

There are also interacting dark energy models \cite{Kumar:2016zpg,DiValentino:2017iww,Kumar:2017dnp,Gomez-Valent:2020mqn,Lucca:2020zjb,vandeBruck:2017idm,Yang:2018euj,Yang:2018uae,Yang:2019uzo,Martinelli:2019dau,DiValentino:2019ffd,DiValentino:2019jae,Pettorino:2013oxa,Yang:2020uga,Yang:2019uog,Yang:2018ubt}, in which dark matter and dark energy have an extra non-gravitational interaction in hope for alleviating this tension. However, there remains some doubts and questions about the detection of dark energy dark matter interaction \cite{DiValentino:2020leo}. Furthermore, decaying dark matter can be a remedy for this problem \cite{Anchordoqui:2015lqa,Berezhiani:2015yta,Vattis:2019efj}. In addition, one can have a higher $H_0$ at the price of extra relativistic degrees of freedom, parameterized by the $N_{\rm eff}$ \cite{Carneiro:2018xwq,Gelmini:2019deq}.

We have also proposed a model to address the $H_0$ tension based on the idea of critical phenomena in \cite{Banihashemi:2018has,Banihashemi:2018oxo,Farhang:2020sij}. In these works we considered that the dark energy experienced a phase transition in its history. This assumption can be natural due to the discrepancy between early and late cosmology.  In \cite{Banihashemi:2018has}, this phase transition has been modeled phenomenologically by a $\tanh$ function. This behavior for dark energy has been assumed independently in \cite{Li:2019yem,Pan:2019hac,Li:2020ybr,Rezaei:2020mrj} as Phenomenological Emergent Dark Energy (PEDE). In this work we would like to generalize our idea in \cite{Banihashemi:2018has} by assigning a more realistic time evolution to dark energy. In the physics of critical phenomena, the Ising model is one of the well-considered models and shares its behavior with many other ones. We will assume that dark energy behaves like an Ising model which results in a very specific time evolution for it.

\section{Critically Emergent Dark Energy (CEDE)}

Adapted from the literature of critical phenomena, we suppose that the dark energy consists of a sort of self interacting micro-structures which carry a local ``order parameter". Here, we attribute a scalar field, $\phi(\boldsymbol{r},z)$, to be representative of this order parameter in each point of space and time (redshift). In fact, this field corresponds to the concept of the coarse grained local magnetization within the Ising model context. We assume the spatial average of $\phi$ is responsible for the background density of the dark energy:
\begin{equation}
\Omega_{\rm DE}(z)=\langle\phi(\boldsymbol{r},z)\rangle\equiv M(z).\label{ave}
\end{equation}

For now, we assume that our dark energy is somehow in thermal contact\footnote{It is important to emphasize that this interaction should be very tiny. Otherwise we expect modification in the behaviour of photons which are very well-constrained. On the other hand, theoretically one can assume that the phase transition in the behaviour of DE is realized by changing of the temperature due to the expansion rate of the universe. In this scenario we could assume $T_{DE}\propto (1+z)^\beta$. The $\beta$ dependence does not change the general behavior of our model and here we assume $\beta=1$. \label{footnote-1}} with the radiation and has a temperature proportional to the cosmological redshift, $T_{\rm DE}\propto 1+z$. At early times or equivalently high temperatures, the local order parameter changes rapidly and there is no long range correlation in the dark energy field. it is completely disordered and its spatial average vanishes. As the universe cools down, order begins to emerge and the field starts to show a long range correlation. After a critical temperature, $T_c$, the average of $\phi$ takes distance from zero and dark energy appears in a continuous phase transition. To formulate this scenario, we use the mathematical description of phase transitions, developed by Ginzburg and Landau \cite{Ginzburg:1950sr}. For this goal, we introduce our effective free energy\footnote{In the literature of critical phenomena it is shown by $L$. But here we used $E_F$ to not make any confusion with the Lagrangian.}
\begin{equation}
	E_F=\int d^3\boldsymbol{r}\bigg[\frac{\gamma}{2}\,(\nabla\phi)^2+m\,t\,\phi^2+\frac{1}{2}\lambda\,\phi^4\bigg],\label{LFE}
\end{equation} 
and demand that the $E_F$ be always stationary with respect to the variations of $\phi$. In the above expression, $t$ is the reduced temperature, $t\equiv\frac{T-T_c}{T_c}$, and $\gamma$, $m$ and $\lambda$ are some positive constants that are supposed to be determined from observations. As it is evident in (\ref{LFE}), we have supposed $\mathbb{Z}_2$ symmetry for our effective free energy; i.e. there isn't any odd power of $\phi$. This symmetry implies that the phase transition is continuous or second order. For the equation of motion of $\phi$ we obtain:
\begin{eqnarray}
-\gamma\,\nabla^2\phi(\boldsymbol{r})+2\,m\,t\,\phi(\boldsymbol{r})+2\,\lambda\,\phi^3(\boldsymbol{r})=0.
\end{eqnarray}
At the first approximation, we consider $\phi$ within the mean field approach: $\phi(\boldsymbol{r})=\langle\phi(\boldsymbol{r})\rangle\equiv M$. Hence it should obey the following equation:
\begin{equation}
2\,m\,t\,M+2\,\lambda\, M^3=0,
\end{equation}
which yields:
\begin{equation}
M=0\quad\text{or}\quad M=\pm\sqrt{-\frac{mt}{\lambda}}=\pm\sqrt{\frac{m}{\lambda}\,\frac{z_c-z}{1+z_c}}.\label{M}
\end{equation}
The last equality comes from our assumption for the dark energy temperature, $T_{DE}\propto 1+z$. The case $M=0$ corresponds to $T>T_c$ and the two other cases belong to  $T<T_c$; after spontaneous symmetry breaking, $M$ takes one of these two possible temporal functionalities\footnote{Note that what is observable is $m/\lambda$. According to (\ref{LFE}), the parameter $m$ encodes the interaction strength between DE and other fields (e.g. photons).  This means even by setting $m$ to be very small, we expect we can have the phase transition but ignore the effects on the other fields. In this scenario we can set $\lambda$ to get an appropriate value for dark energy density.}. We choose $M$ to take the positive one, because negative values for the dark energy density are not very pleasing philosophically. It is worth to clarify that we assumed quasi-static case for the DE system. This effectively means the interaction has smaller time scale in comparison to the Hubble time.

\begin{figure}
	\centering
	\includegraphics[width=\linewidth]{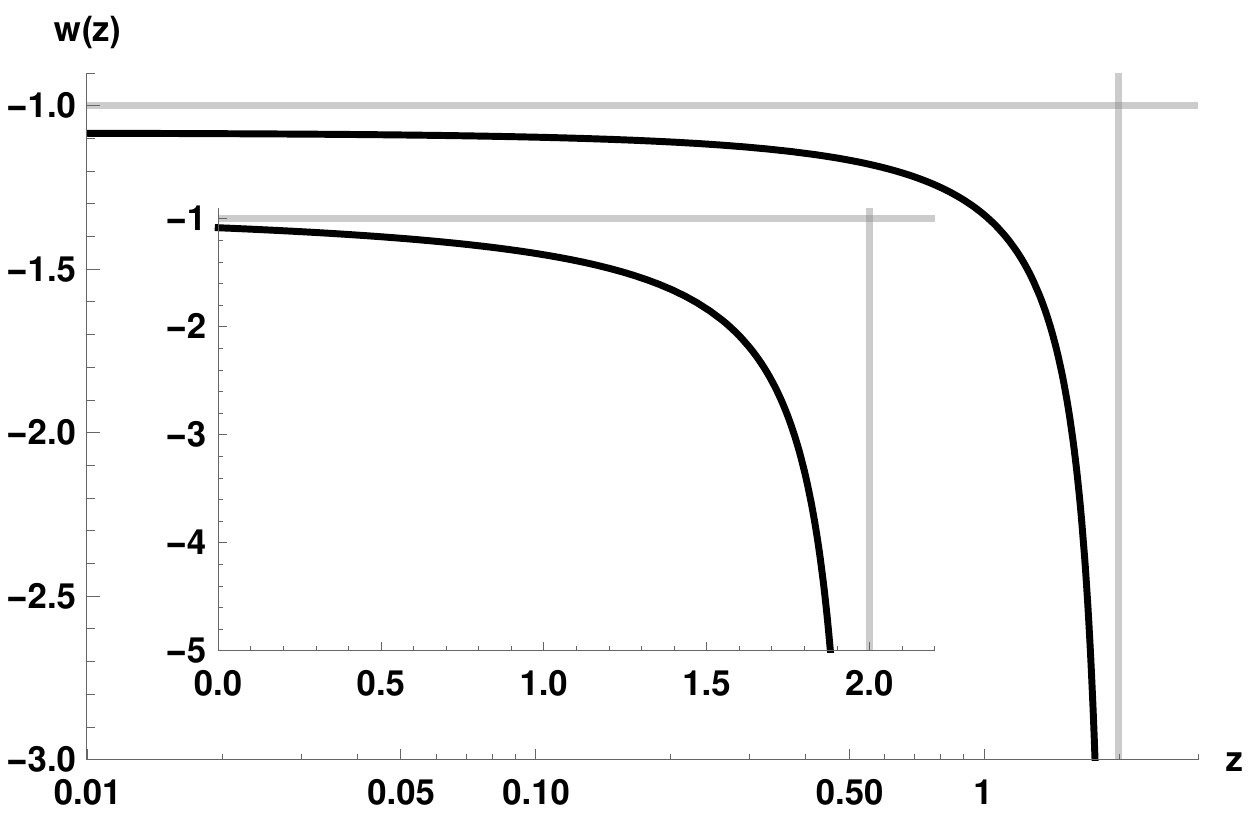}
	\caption{The equation of state of dark energy versus redshift for a typical transition redshift, $z_c=2$. The grey lines are horizontal and vertical asymptotes of this function which lie at $w=-1$ and $z=z_c$, respectively. Note that as we discussed in the draft, $\rho_{\rm DE}(z)$ is the physical quantity and well behaved at $z_c$ even if $w_{\rm DE}(z)$ diverges at the critical redshift. The sub-plot inside, is the same function with linear horizontal axis.}
	\label{fig:wz}
\end{figure}

 In addition, spatially flatness of the universe fixes the coefficient $m/\lambda=(1-\Omega_m-\Omega_r)^2\,(1+z_c)/z_c$ in (\ref{M}) and we obtain
\begin{equation}\label{ODE}
\Omega_{\rm DE}(z)=(1-\Omega_m-\Omega_r)\sqrt{\frac{z_c-z}{z_c}},
\end{equation}
where $\Omega_m$ and $\Omega_r$ are the matter and radiation fractional densities at present time, respectively. So in this step we have the form of the Friedmann equation,
\begin{equation}
 H^2=H_0^2\,\bigg[\Omega_m(1+z)^3+\Omega_r(1+z)^4+\Omega_{\rm DE}(z)\bigg].
\end{equation}
We can deduce the dark energy's equation of state from the continuity equation, namely,
\begin{equation}\label{Ceq}
\rho'_{\rm DE}(z)-\,\frac{3}{1+z}\,\rho_{\rm DE}(z)\,\big[1+w_{\rm DE}(z)\big]=0,
\end{equation}
where $'$ denotes the derivative with respect to the redshift. By substituting $\rho_{\rm DE}(z)$ from (\ref{ODE}) into (\ref{Ceq}) we can read  $w_{\rm DE}(z)$ as
\begin{equation}
w_{\rm DE}(z)=-1-\frac{1+z}{6(z_c-z)}.
\end{equation}
This relation holds for any moment after onset of the phase transition and before that time, since the density of dark energy is zero, $w_{\rm DE}$ is not well-defined. Right at the moment of transition, $w_{\rm DE}$ diverges toward $-\infty$ and afterwards, at far future, approaches $-1$ (c.f. figure \ref{fig:wz}). Physically, this means the DE effectively is in a phantom phase and dynamically approaches to the cosmological constant. This is a behavior that we expect from emergent dark energy models \cite{Li:2019yem,Li:2020ybr}.  We emphasize that since $w_{\rm DE}$ is not appeared in the physics directly then $w_{\rm DE}=-\infty$ does not give any physically divergences in our model\footnote{Note that the physics is given by $\rho(a)=\rho_0\times exp[-3\int_{a_0}^{a}(1+w(a))d\ln a]$ where an infinite $w_{\rm DE}$ at one point cannot make any divergences.}. It is worth to add that we could expect this behaviour since at the critical temperature, it is a very natural behaviour in the physics of critical phenomena to see a discontinuity in some of the parameters.

\section{Analysis}

The parameter space we want to put constraint on, is
\begin{equation}
\mathcal{P}=\{\Omega_bh^2,\Omega_ch^2,100\Theta_{MC},\tau,n_s,\ln[10^{10}A_s],a_c\},
\end{equation} 
which shows that our model has six parameters in common with $\Lambda$CDM and an extra free parameter $a_c$ which represents the scale factor of transition.

In order to put constraint on these free parameters, we focus on the following two data sets:
\begin{itemize}
	\item Planck 2018 temperature and polarization angular power spectra, i.e. combination of the \texttt{\textbf{Commander, SimALL}} and \texttt{\textbf{plikTT,TE,EE}} likelihoods.\cite{Aghanim:2019ame}. We refer to this data as \textbf{CMB}.

	\item The latest measurement of $H_0$ by Riess \textit{et al.} \cite{Riess:2019cxk}. We refer to this data point as \textbf{R19}.
\end{itemize}  
In this work we do not add the other datasets e.g. BAO or Supernovae. The first reason is that we want to see if our model can reconcile between CMB and R19. The second reason is that it is not clear if BAO and SNe datasets are consistent with CMB. We will study this consistency in our future works. We make use of publicly available code \texttt{CAMB} \cite{camb,Lewis:1999bs}, to calculate the predictions of CEDE for the observables described above. For sampling the parameter space, we use \texttt{CosmoMC} \cite{Lewis:2002ah,Lewis:2013hha}. We use \texttt{GetDist} \cite{Lewis:2019xzd} in order to extract the parameter's posteriors from the MCMC chains and to plot likelihood contours.

\begin{figure}
	\centering
	\includegraphics[width=\linewidth]{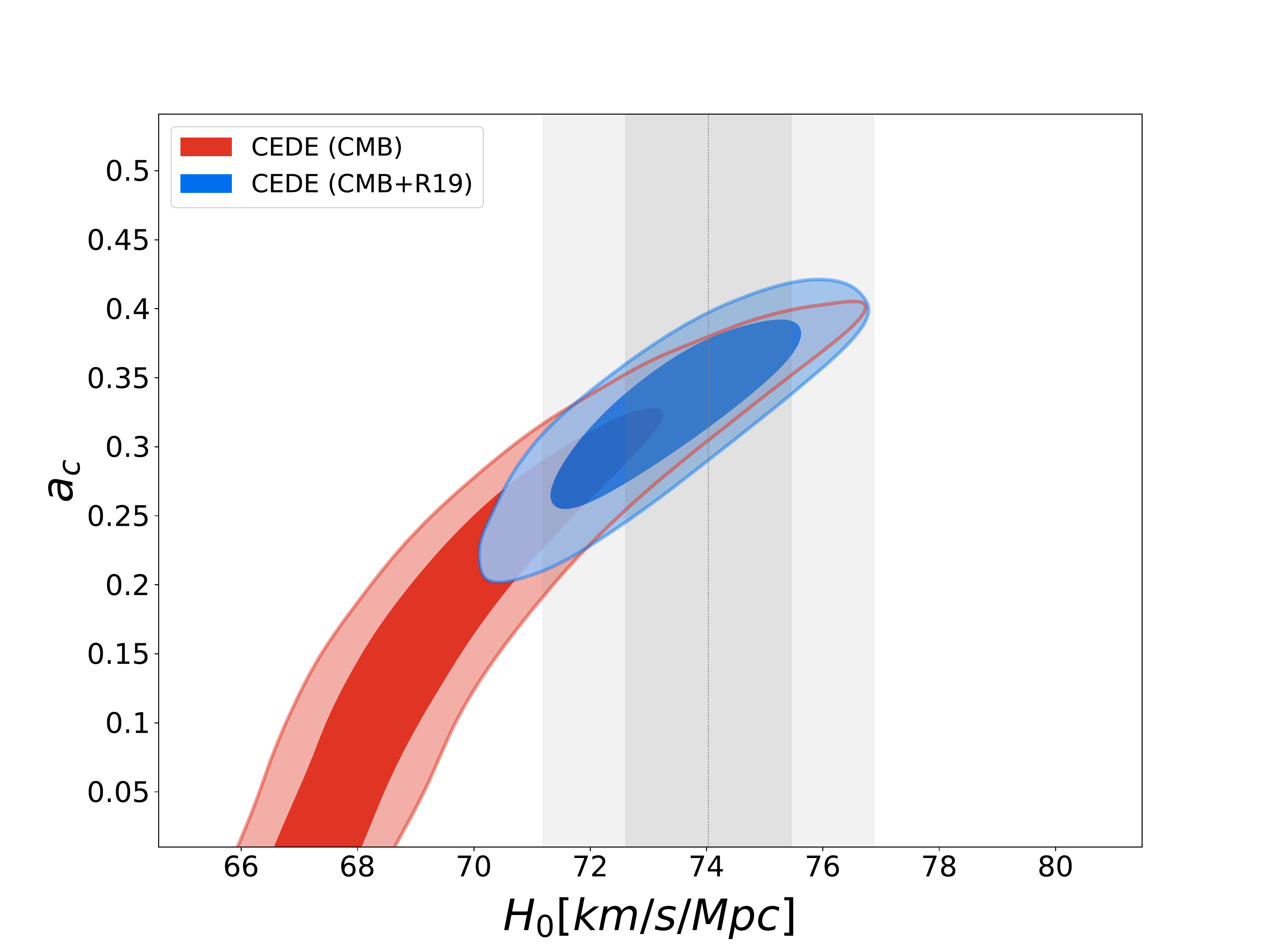}
	\caption{The $1-2\sigma$ contours for $a_c$ vs $H_0$ is plotted for CEDE model where Planck \textbf{CMB} data is used without and with \textbf{R19} data, in red and in blue, respectively. The grey bands show the $1-2\sigma$ values for $H_0$ as reported by Riess et al. from local measurments. The red contours show that CEDE model has no inconsistency with \textbf{R19} which means we can combine \textbf{R19} and \textbf{CMB} data together to get the joint constraints (blue contour). It is worth to mention that CEDE reduces to $\Lambda$CDM when the transition occurs very early (effectively at $a_c$ tends to $0$). The red contours show this property very clearly: for very small $a_c$ the $H_0$ value becomes closer to what Planck reports for $\Lambda$CDM.}
	\label{fig:h0-at}
\end{figure}

\begin{figure}
	\centering
	\includegraphics[width=\linewidth]{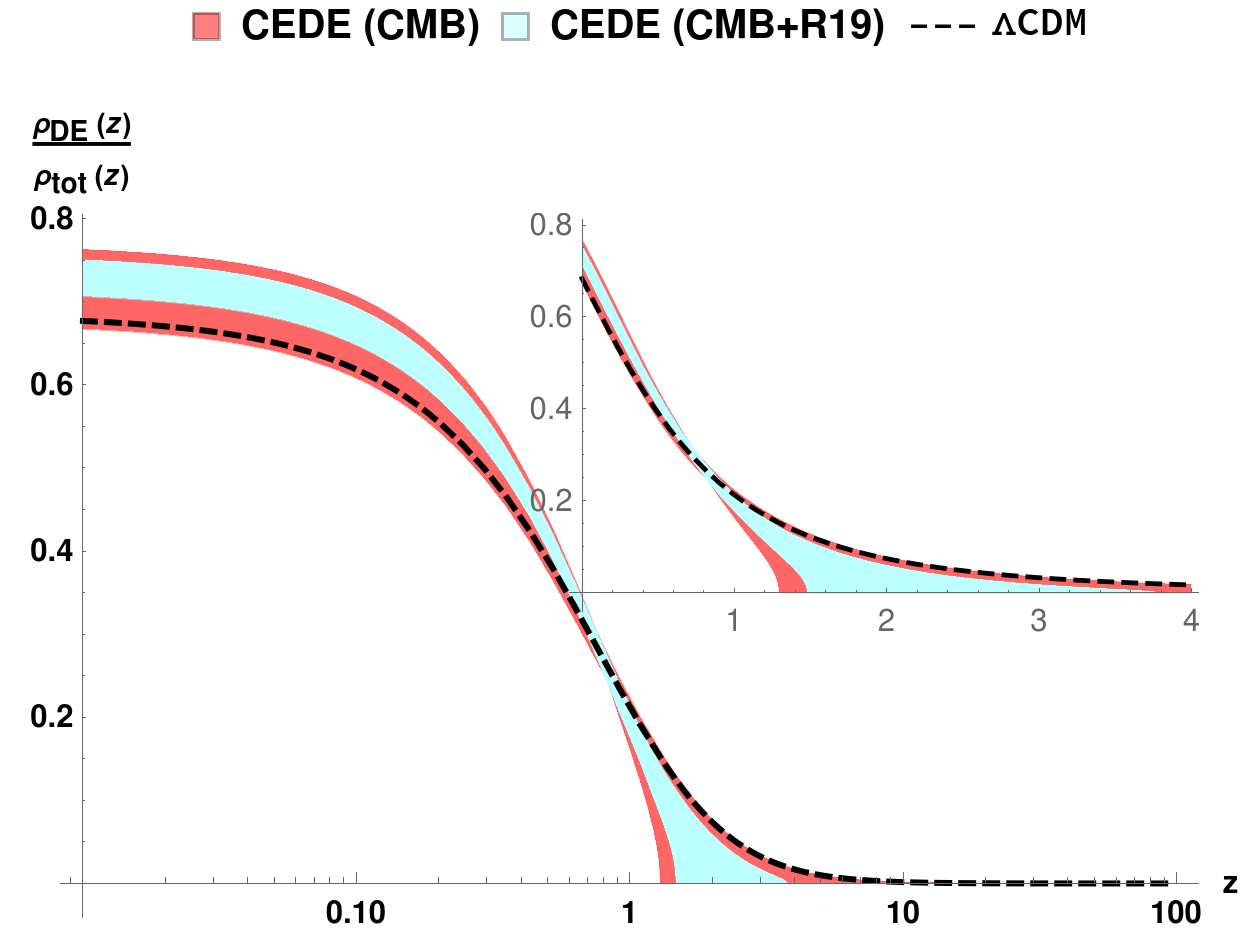}
	\caption{The fractional density of dark energy is plotted versus the redshift. The dashed black line represents this function for the $\Lambda$CDM model best fit values. The red line region shows the same for $1\sigma$ values of CEDE when it is constrained by only \textbf{CMB}. Obviously, for this case, CEDE includes $\Lambda$CDM as we discussed in the paper. But including \textbf{R19} excludes the $\Lambda$CDM as it is shown by $1\sigma$ region in blue. While the main plot is logarithmic in redshift, the interior sub-plot is linear up to $z=4$.}
	\label{fig:fractionalcmb}
\end{figure}

\section{Results}
\begin{figure*}
	\centering
	\includegraphics[width=\linewidth]{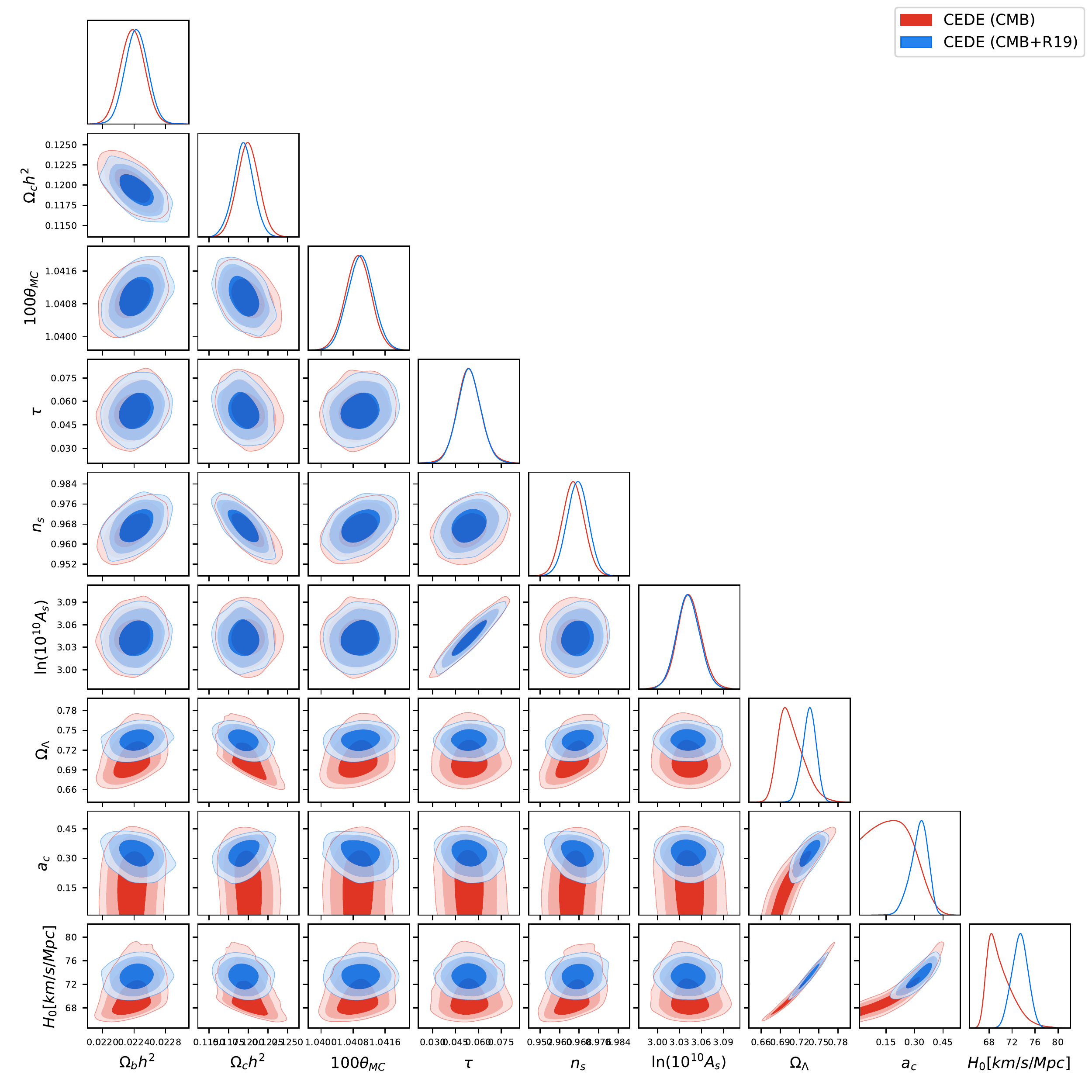}
	\caption{$68\%$, $95\%$ and $99\%$ parameter constraint contours for CEDE  from two sets of data. Note that blue contours are narrower because of inclusion of the \textbf{R19}; but still over lapping with the red contours.}
	\label{fig:triangle}
\end{figure*}

The first very important step is to see if independent datasets are compatible for a theoretical setup or not. If they are not compatible then we are not allowed to use both of them at the same time (that usually would result to tight constraints for the model parameters). In the framework of $\Lambda$CDM, \textbf{CMB} and \textbf{R19} are more than $4\sigma$ incompatible which is the Hubble tension. To check the compatibility between these two datasets we have put constraints on CEDE parameters just with \textbf{CMB} dataset. The main result can be seen in figure \ref{fig:h0-at} where the red likelihood shows the $2-d$ likelihood for $H_0$ and $a_c$ in CEDE. The result is very promising: constraints from \textbf{CMB} data has no tension with \textbf{R19} data in this model. Obviously, the $1\sigma$ likelihood touches \textbf{R19}'s $1\sigma$; and $2\sigma$ likelhood for $H_0$ in CEDE reaches to upper $2\sigma$ value for \textbf{R19}. In figure \ref{fig:triangle}, we have shown the $2-d$ and $1-d$ likelihoods for our free parameters as well as derived $H_0$ and $\Omega_\Lambda$. It is obvious from the last row that \textbf{R19} does not show any inconsistencies in our free parameters. Consequently, we can add up \textbf{CMB} and \textbf{R19} datasets. The likelihoods for CEDE has been shown in figures \ref{fig:h0-at} and \ref{fig:triangle} in blue. The best fit values are reported in table \ref{tab:best-fit} for $\Lambda$CDM and CEDE in the presence of \textbf{CMB} with and without \textbf{R19}. Note that we have reported the best values for $\Lambda$CDM parameters constrained by \textbf{CMB} and \textbf{R19} but it can not be trusted since these two datasets are incompatible in the framework of $\Lambda$CDM.

The main result is that including \textbf{R19} needs a non-trivial phase transition in DE by excluding lower values for transition scale factor $a_c$. This lower value at $2\sigma$ is around $a_c\sim 0.2$ which is equivalent to $z_c\sim 4$.  To explain it let us emphasize that for very small values of $a_c$ the transition occurs at very high redshifts. This means DE starts to grow from zero to a constant $\Omega_\Lambda$ very early. This transition is so fast and occurs far before DE domination. So it has no observational effects. In other words CEDE behaves exactly same as $\Lambda$CDM for very high transition redshifts. The results show that in the framework of CEDE, $\Lambda$CDM can explain \textbf{CMB} as good as CEDE with non-trivial phase transition. But including \textbf{R19} breaks this degeneracy in favour of a phase transition in the late time behavior of dark energy. To study this fact more clearly we have plotted the dark energy fractional density ($1\sigma$ values) as a function of redshift in figure \ref{fig:fractionalcmb}. It is obvious from this plot that having $\Lambda$CDM is valid in CEDE framework when we use only \textbf{CMB} but including \textbf{R19} excludes it.\\

\begin{table*}
	\begin{scriptsize}
		\begin{tabular}{|c|c|c|c|c|c|}
			
			\hline&&&&&\\
			$\quad$Parameter$\quad$&$\quad$Prior$\quad$&$\Lambda$CDM (\textbf{CMB})&$\Lambda$CDM (\textbf{CMB+R19})&$\quad$ CEDE (\textbf{CMB})$\quad$&$\quad$ CEDE (\textbf{CMB+R19})$\quad$\\
			\hline&&&&&\\
			$\Omega_bh^2$&[0.005\ ,\ 0.1]&$0.02236\pm 0.00015$&$0.02254\pm 0.00014$&$0.02243^{+0.00013}_{-0.00014}$&$0.02243\pm 0.00014$\\&&&&&\\
			
			$\Omega_ch^2$&[0.001\ ,\ 0.99]&$0.1202\pm 0.0014$&$0.1179\pm 0.0012$&$ 0.1199\pm 0.0014$&$0.1194\pm 0.0013$\\&&&&&\\
			
			100$\Theta_{MC}$&[0.5\ ,\ 10]&$1.04091\pm 0.00032$&$1.04120\pm 0.00030$&$1.04093\pm 0.00032$&$1.04098\pm 0.00032$\\&&&&&\\
			
			$\tau$&[0.01\ ,\ 0.8]&$0.0545\pm 0.0077$&$0.0579\pm 0.0084$&$0.0539\pm 0.0080$&$ 0.0539\pm 0.0077$\\&&&&&\\
			
			$n_s$&[0.8\ ,\ 1.2]&$0.9648\pm 0.0043$&$0.9703\pm 0.0042$&$0.9656\pm 0.0045$&$ 0.9674\pm 0.0043$\\&&&&&\\
			
			$\ln[10^{10}A_s]$&[2\ ,\ 4]&$3.045^{+0.014}_{-0.016}$&$3.047\pm 0.017$&$3.043\pm 0.017$&$3.042\pm 0.016$\\&&&&&\\
			
			$a_c$&[0.01\ ,\ 1]&-&-&$ 0.185^{+0.087}_{-0.14}$&$0.325^{+0.052}_{-0.036}$\\&&&&&\\

			$\Omega_{\Lambda}$&-&$0.6837\pm 0.0084$&$0.6976\pm 0.0072$&$0.707^{+0.014}_{-0.023}$&$0.735^{+0.011}_{-0.0097}$\\&&&&&\\
			
			$H_0\ [\rm km/s/Mpc]$&-&$67.29\pm 0.60$&$68.32\pm 0.55$&$70.0^{+1.2}_{-2.7}$&$73.4\pm 1.4$\\\hline&&&&&\\
			
			Total $\chi^2_{\rm min}$&-&$\quad2772.3\quad$&$\quad\boldsymbol{2792.9}\quad$&$\quad2771.7\quad$&$\quad\boldsymbol{2775.0}\quad$\\&&&&&\\
			
			$\chi^2_{\rm\textbf{CMB}}$&-&$2768.9$&$2769.6$&$2769.6$&$2769.5$\\&&&&&\\
			
			$\chi^2_{\rm\textbf{R19}}$&-&-&20.4&-&1.0\\
			\hline
		\end{tabular}	
	\end{scriptsize}
	
	\caption{\label{tab:best-fit}The best fit values and $68\%$ CL intervals for $\Lambda$CDM and CEDE parameters when two combinations of data are used. When only \textbf{CMB} is used, CEDE shows a slightly lower $\chi^2$ with respect to $\Lambda$CDM and inclusion of \textbf{R19} makes our $\Delta\chi^2$ much more negative and hence preferred.} \label{back-chi2}
\end{table*}

\section{Discussions}
In this work we have shown a specific form of transition in the behaviour of dark energy can be a very promising framework to address the Hubble tension. This specific from of transition is based on the physics of critical phenomena, specifically the Ising model. We named our model ``critically emergent dark energy", CEDE. As it is shown figure \ref{fig:h0-at}, in CEDE there is no tension between CMB from Planck and local $H_0$ measurement by Riess et al. Consequently we can add up these datasets without any worries.

Theoretically, in the critical phenomena literature, the phase transition in the behaviour of a system means there is a substructure for that system. So we think the Hubble tension can be a hint for existence of substructure of dark energy. This has a very rich phenomenology observationally: What are the effects of these substructures?  e.g. in ISW signal or structure formation? Do they make dark energy to be clustered? Another very interesting question is that what happened at the redshift of transition: according to physics of critical phenomena at the critical point, we expect the correlation length become infinity. This may cause a very specific fingerprint in the cosmological observables at the redshift of transition. 

Here, we checked CEDE only against  \textbf{CMB} and \textbf{R19} to show the main properties of this model but we have a plan to check it against the other datasets including BAO and SNe Ia compilations. Another very interesting path to follow is comparing CEDE with PEDE \cite{Li:2019yem} and GEDE \cite{Li:2020ybr} phenomenological models of emergent dark energy as they have effectively very similar properties.

\section*{Acknowledgments}
We thank E. Linder and A. Starobinsky for very useful comments on the draft. AB and NK thank Iran National Science Foundation (INSF) for supporting this work partly under project no. 98022568.

\end{document}